# Finite element analysis of ion-implanted diamond surface swelling

Federico Bosia [a,f§], Silvia Calusi [b], Lorenzo Giuntini [c], Stefano Lagomarsino [d], Alessandro Lo Giudice [b,f], Mirko Massi [c], Paolo Olivero [b,f], Federico Picollo [b,f], Silvio Sciortino [e], Andrea Sordini [d], Maurizio Vannoni [d], Ettore Vittone [b,f]

[a] Department of Theoretical Physics, University of Torino, Italy

[b] Department of Experimental Physics and "Nanostructured Interfaces and Surfaces" Centre of Excellence, University of Torino, Italy

[c] Department of Physics, University of Firenze, Italy

[d] CNR Istituto Nazionale di Ottica Applicata (INOA), Firenze, Italy

[e] Department of Energetics, University of Firenze, Italy

[f] INFN Sezione di Torino, University of Torino, Italy

**Abstract**

We present experimental results and numerical Finite Element analysis to describe surface swelling due to the creation of buried graphite-like inclusions in diamond substrates subjected to MeV ion implantation. Numerical predictions are compared to experimental data for MeV proton and helium implantations, performed with scanning ion microbeams. Swelling values are measured with white-light interferometric profilometry in both cases. Simulations are based on a model which accounts for the through-the-thickness variation of mechanical parameters in the material, as a function of ion type, fluence and energy. Surface deformation profiles and internal stress distributions are analyzed and numerical results are seen to adequately fit experimental data. Results allow us to draw conclusions on structural damage mechanisms in diamond for different MeV ion implantations.

**Keywords**

Diamond; Ion damage; Numerical analysis

---

[§] Corresponding author: Federico Bosia, fbosia@to.infn.it

# 1. Introduction

Diamond is a material of great interest for its extreme physical and chemical properties: high hardness and Young modulus, wide transparency band, chemical inertness, full bio-compatibility, etc. The implantation of high energy (MeV) ions allows the fabrication and functionalization of this material, because of its peculiar characteristic of converting the pristine crystal to significantly different structural phases (graphite, amorphous and glassy carbon) when its lattice structure is critically damaged. The strongly non-uniform damage profile of MeV ions allows the direct creation of specific regions of the material with different physical properties (i.e. electrical conductivity [1; 2], refractive index [3; 4], etc.) or different reactivity to subsequent processing (i.e. selective chemical etching of sacrificial layers with respect to the chemically inert diamond matrix) [5]. All of this can be exploited to fabricate a range of micro-devices, ranging from bio-sensors to micro-electromechanical systems (MEMS) and optical devices [6].

Although the role of implantation fluence has been investigated in several works [7; 8; 9; 10; 11] some uncertainty remains on the structural modifications occurring in diamond as a function of other relevant parameters, namely the impinging ion type, energy, implantation temperature, annealing temperature, local stress, etc [12]. With regards to ion fluence, a critical damage level $D_C$ has been identified in the literature above which diamond is subject to permanent amorphization and subsequent graphitization upon thermal annealing, but this value seems to depend on the depth of the damaged layer, although no specific dependence has been established [9; 13; 14; 15].

One relevant consequence of the structural modifications due to ion implantation is a density variation in the damaged diamond, i.e. a constrained volume expansion which leads to surface swelling in correspondence with the irradiated region [7; 8; 9; 10; 11]. This mechanical effect can be exploited to deduce information regarding the structural modifications occurring in the substrate.

In the present work, we compare theoretically predicted swelling values, obtained by adopting a simple phenomenological model that uses the critical damage level as a parameter, to the experimentally measured values. The analysis is carried out using Finite Element Model (FEM) simulations, in order to correctly take into account the complex stress state and the related deformations.

The paper is structured as follows: in Section 2, the model for surface swelling is outlined; in Section 3 the experimental procedure and the relevant results are described; in Section 4 the FEM calculations are presented and a comparison between experimental and numerical data is given.

## 2. Surface swelling in ion-implanted diamond

A simple phenomenological model was developed that correctly describes the density variation in diamond due to modifications in the crystal lattice during ion implantation. With the "SRIM - The Stopping and Range of Ions in Matter" (SRIM) Monte Carlo simulation code [16] it is possible to estimate the profile of the linear density of vacancies $\lambda(z)$, if no saturation effects are taken into account (see below). Fig. 1a shows $\lambda(z)$ (expressed in number of vacancies per unit of length in the depth direction, per incoming ion) in the three cases considered in this work: 1.8 MeV He, 2 MeV H and 3 MeV ions. It is apparent that a large fraction of the nuclear energy loss (which is responsible for structural damage) occurs at end of range of ions, i.e. at depths of about 3 $\mu$m, 25 $\mu$m and 50 $\mu$m, respectively.

It is expected that the formation of vacancies and interstitials leads to an amorphization of the material, i.e. a transition from the density of diamond ($\rho_d$ = 3.515 g·cm$^{-3}$) to that of amorphous carbon ($\rho_{aC}$ ≈ 1.557 g·cm$^{-3}$) [17]. The greater the damage density, the smaller the density of the damaged diamond $\rho$ should become, with saturation towards the lower bound value of $\rho_{aC}$. As mentioned above, this saturation is not predicted by SRIM Monte Carlo code, in which the interaction of each single ion with the pristine crystalline structure is simulated and no cumulative damage effects are taken into account. The saturation effect can be derived from the assumption that the recombination probability for a vacancy in a damage cascade is proportional to the vacancy density $\rho_V$ itself: $P_{REC} = \rho_V/\alpha$ [18], where parameter $\alpha$ expresses the critical vacancy density at which all additional vacancies recombine with existing interstitial atoms in the damaged material. We can therefore write the differential increment in vacancy density $d\rho_V$ due to an increment in fluence $dF$ for an ion implantation characterized by damage profile $\lambda(z)$:

$$d\rho_V(F,z) = [1 - P_{REC}(z)] \cdot \lambda(z) \cdot dF = \left[1 - \frac{\rho_V(F,z)}{\alpha}\right] \cdot \lambda(z) \cdot dF \qquad (1)$$

By integrating eq. (1) with the boundary condition $\rho_V(F=0,z)=0$, we obtain the expression of the induced vacancy density at a given depth for a given implantation fluence, which takes into account the above-mentioned saturation effect:

$$\rho_V(F,z) = \alpha \cdot \left[1 - \exp\left(-\frac{\lambda(z) \cdot F}{\alpha}\right)\right] \qquad (2)$$

From eq. (2) we infer that the vacancy density asymptotically approaches the critical density $\alpha$, which is therefore bound to be smaller than the atomic density of the pristine diamond structure $\rho_d$. It is worth stressing that $\alpha$ defines the slope of the exponential growth of the vacancy density towards the saturation value.

We further generalize the model in [18] by assuming that the density of the damaged material is determined by the vacancy density by:

$$\rho(F,z) = \rho_d - \beta \cdot \rho_V(F,z) = \rho_d - \beta \cdot \alpha \cdot \left[1 - \exp\left(-\frac{\lambda(z) \cdot F}{\alpha}\right)\right] \quad (3)$$

where the empirical parameter $\beta$ is adopted in a simple linear model to quantify the effect of the induced vacancy density on the density of the damaged material. By applying the boundary condition that sets the saturation density of the damaged material to the density of amorphous carbon $\rho_{aC}$, we obtain $\beta = (\rho_d - \rho_{aC})/\alpha$, hence the relation:

$$\rho(F,z) = \rho_d - (\rho_d - \rho_{aC}) \cdot \left(1 - e^{-\frac{F\lambda(z)}{\alpha}}\right) \quad (4)$$

In our model, the contribution to the material density due to the accommodation of implanted ions is neglected. Such an assumption is acceptable in the case of light ions such as those used in this work (H and He).

The parameter $\alpha$ is reminiscent of the critical damage level $D_C$ [14], except that it is relative to samples *before* annealing: following from this analogy, when the implantation fluence is high enough (i.e. $F \gg \alpha/\lambda(z)$), the damaged material is envisaged to be subjected to a permanent amorphization process. $\alpha$ will be considered as a free parameter in the following simulations, and will be evaluated for any ion type and energy, by fitting the numerical evaluation of swelling with the experimental values as function of different implantation fluences.

The expected density variation as a function of depth $z$ and fluence $F$ is shown Fig. 1b in the case of 1.8 MeV He ions in the fluence range $5 \cdot 10^{15} - 5 \cdot 10^{17}$ cm$^{-2}$, for $\alpha = 1.1 \cdot 10^{23}$ cm$^{-3}$. The damaged layer in this case extends to a depth of about 3.3 µm. The saturation to the amorphous carbon density value $\rho_{aC}$ is apparent. The swelling data relative to this case will be discussed in the experimental and numerical sections.

Futhermore, we assume that the variation of the other mechanical parameters of the damaged substrate, i.e. Young's modulus $E(F, z)$ and Poisson's ratio $\nu(F, z)$, also follow similar functional forms as that in eq. (4). This is equivalent to adopting a sort of "rule of mixtures" description for the damaged diamond, where all material properties vary smoothly between those of diamond and those of amorphous carbon.

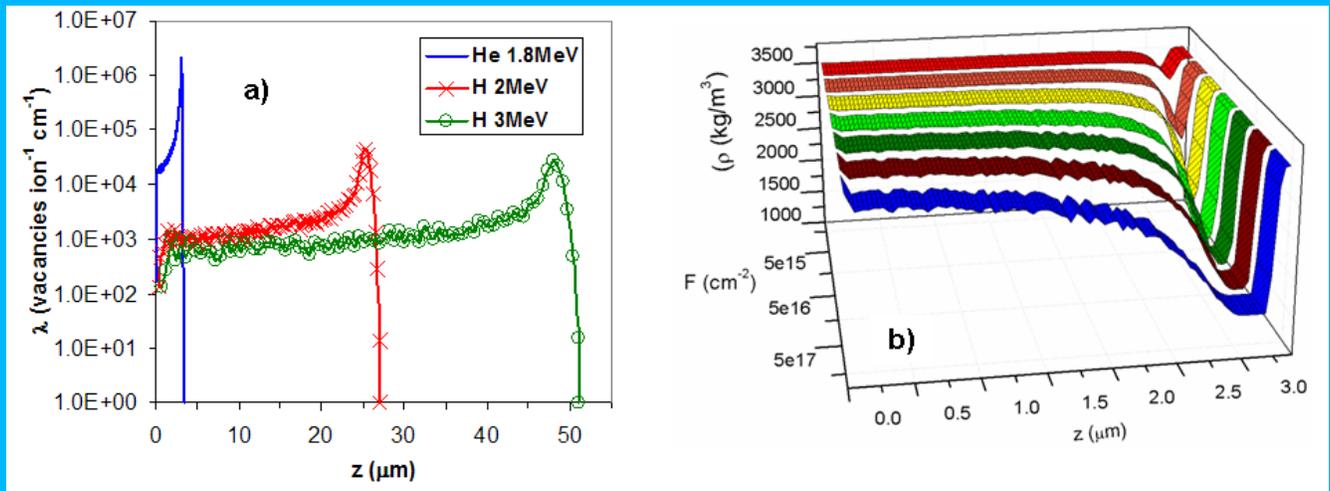

Figure 1. *a)* depth profile of the vacancy linear density $\lambda(z)$ for He 1.8 MeV, H 2 MeV, and H 3 MeV ion implantations; *b)* damaged diamond density variation as a function of depth $z$ and fluence $F$ for a 1.8 MeV He implantation ($\alpha=1.1\cdot10^{23}$ cm$^{-3}$).

## 3. Experimental

### 3.1 Samples

Ion implantation was performed on various artificial single crystal diamonds that were produced with different synthesis techniques. Firstly, samples grown with the "high pressure high temperature" (HPHT) method by Sumitomo were employed. Such samples are classified as type Ib, indicating a substitutional nitrogen concentration comprised between 10 and 100 ppm. The samples are cut along the 100 crystal direction and usually consist of different growth sectors. Secondly, samples grown with chemical vapor deposition (CVD) technique by ElementSix were used. Such samples are characterized by a higher purity and are classified as type IIa, having a nitrogen concentration below 0.1 ppm. The crystals consist of a single 100 growth sector. In both cases, the size is $3\times3\times1.5$ mm$^3$, and the two opposite large faces are optically polished. Although characterized by different impurity concentrations, the sample can be reasonably expected to display the same mechanical properties, consisting in both cases in high-quality single-crystals.

**3.2 Ion implantation**

In order to study the damage-induced swelling process in different experimental conditions, the samples were implanted in a broad range of fluences with different ions species and energies. 1.8 MeV MeV He ions were implanted at the ion microbeam line of the INFN Legnaro National Laboratories (Padova), while 2 and 3 MeV H ion implantations were performed at the external microbeam line of the LABEC INFN facility (Firenze). In both cases, the samples were implanted in frontal geometry on their polished surfaces. In order to achieve a uniform fluence delivery in the implantation process, ~125×125 µm$^2$ square areas were implanted by raster scanning an ion beam with size of 20-30 µm. In the former case, the implantation fluence was controlled in real time by monitoring the x-ray yield from a thin metal layer evaporated on the sample surface, while in the latter case the x-ray emission from the beam exit window was employed, after a suitable calibration with a Faraday Cup [19]. The implantations were performed at room temperature, with ion currents of ~1 nA. In these conditions, implantations in the fluence ranges of $10^{15}$-$10^{17}$ cm$^{-2}$ could be performed in times varying from few minutes to ~1 hour.

### 3.3 Profilometry measurements

Surface swelling data were acquired at the Interferometry laboratories of the Istituto Nazionale di Ottica Applicata (INOA) with a Zygo NewView 6000 system, which exploits white light interferometry to provide detailed, non contact measurements of 3-D profiles. A vertical resolution of 0.1 nm was achieved over a lateral range up to 150 µm, while lateral resolution varied from 4.6 µm up to 0.6 µm, depending on the objective. Fig. 2 illustrates an example of a measured swelling profile for a diamond substrate implanted with 1.8 MeV He ions. The implanted area measures approximately 110×170 µm$^2$ and the average swelling in the central region of this area is about 100 nm.

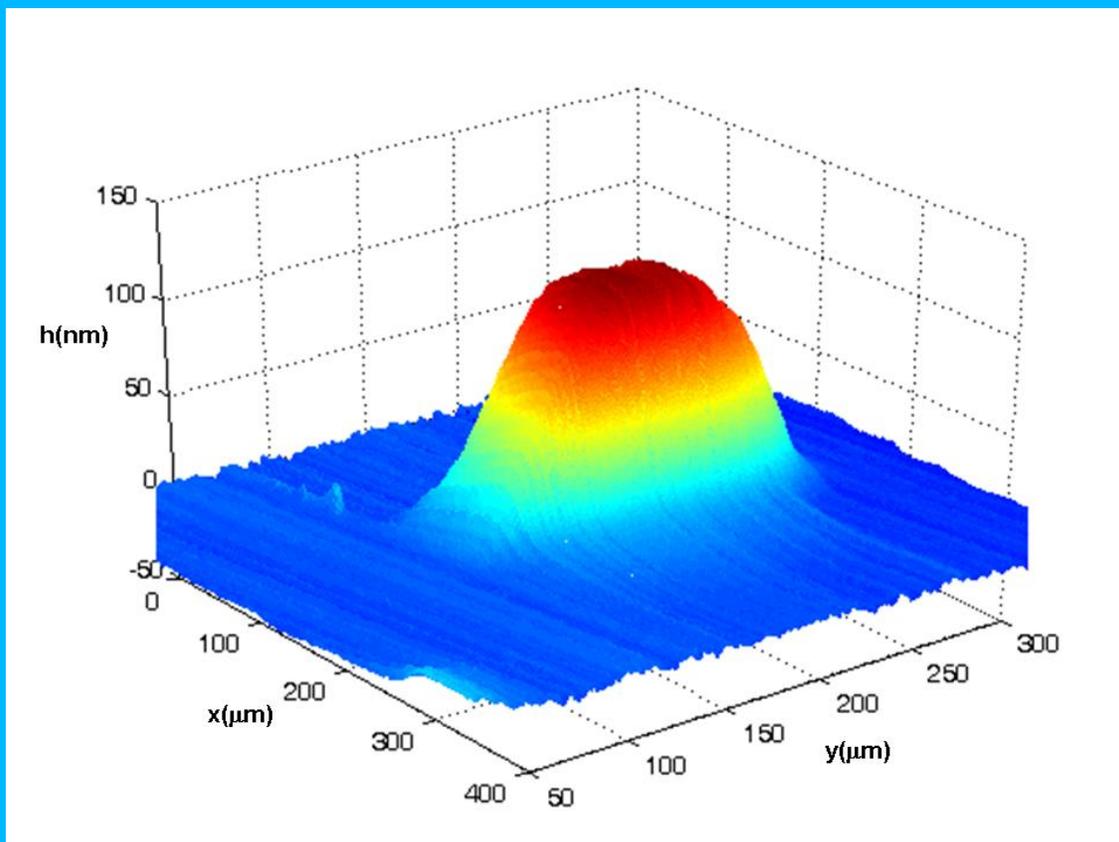

*Figure 2.* Experimentally measured swelling *h* for a 1.8 MeV He implantation ($F$=3.67 10$^{16}$ cm$^{-2}$) using the profilometry technique based on white light interferometry.

## 4. Numerical results

FEM simulations are performed by imposing a constrained isotropic volume expansion in the damaged diamond substrate, which is proportional to the local density variation, as evaluated in the above-mentioned model. This procedure is similar to imposing a thermal expansion, with the local infinitesimal volume variation playing the role of the thermal expansion coefficient.

### 4.1 Finite element model

Simulations were carried out using the commercial software COMSOL Multiphysics, using the "Structural mechanics" module [20]. Specimen geometry is reproduced and meshed both in 2-D and 3-D simulations. The analytical expression of eq. (4) is used, together with the damage profile $\lambda(z)$ resulting from SRIM simulations for 1.8 MeV He, 2 MeV H and 3 MeV H ions. The mechanical properties of diamond and amorphous carbon are $\rho_d$ =3.515 g·cm$^{-3}$, $E_d$ =1144.6 GPa, $\nu_d$ = 0.2, and $\rho_{aC}$ =1.557 g·cm$^{-3}$, $E_g$ = 21.38 GPa, $\nu_g$=0.184, respectively [17]. Some uncertainty remains on the most appropriate value to be chosen for $\rho_{aC}$, however this issue will be addressed in future works. Small uncertainties on the remaining parameters have little influence on the swelling values obtained in simulations. Various simulations were carried out by varying of the critical density $\alpha$, adopting a "best fit" criterion for the resulting swelling with respect to experimental data.

Fig. 3a shows an example of the deformation profile evaluated by FEM calculations relevant to the irradiation conditions Fig. 2 and assuming $\alpha$=1.1·10$^{23}$ cm$^{-3}$. As visible from the comparison of Figs. 2 and 3a, the result of the numeric calculation is in excellent agreement with experimental results. The simulations also allow the evaluation of the stress distributions in the substrate after the internal damaged diamond has expanded. Fig 3b illustrates a typical through-the-thickness Von Mises stress variation for a 3 MeV H implantation ($F$=2·10$^{16}$ cm$^{-2}$, $\alpha$ = 1.65·10$^{23}$ cm$^{-3}$ ).

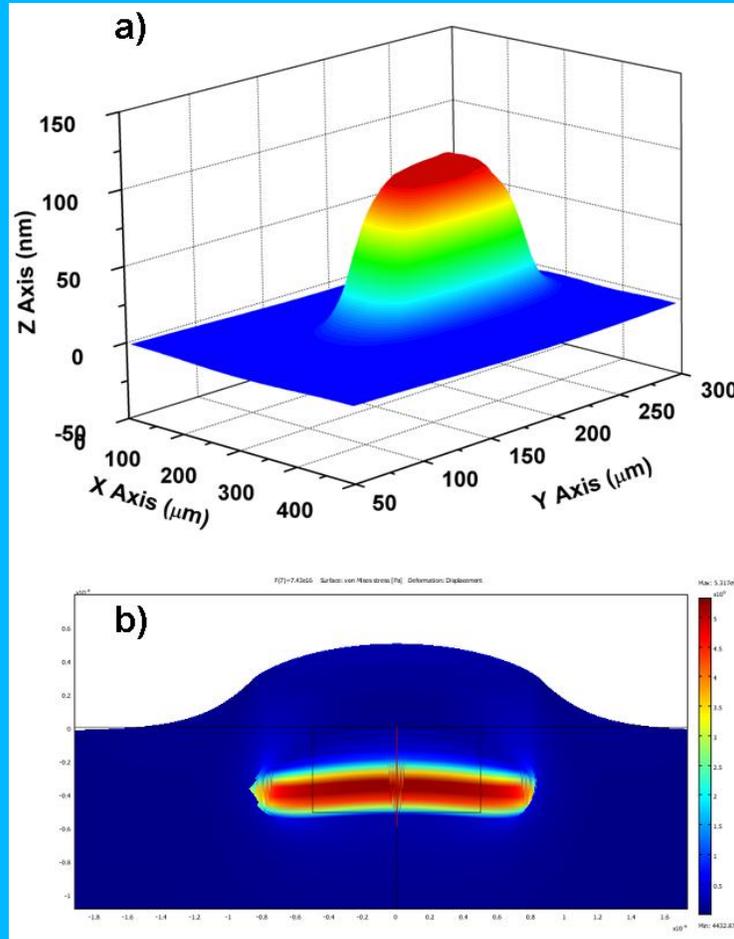

Figure 3: *a)* FEM simulation swelling results for 1.8MeV He implantations with $F=3.67 \cdot 10^{16}$ cm$^{-2}$ and $\alpha=1.1 \cdot 10^{23}$ cm$^{-3}$. *b)* Simulated through-the-thickness Von Mises stress for a 3 MeV H implantation with $F=2 \cdot 10^{16}$ cm$^{-2}$ and $\alpha=1.65 \cdot 10^{23}$ cm$^{-3}$.

**4.2 Comparison with experimental data**

Numerical FEM simulations were carried out for the 3 cases considered, varying the parameter $\alpha$ to obtain the best agreement with experimental data. The "swelling vs fluence" plot in Figs. 4a shows experimental and numerical values for the He implantation at 1.8 MeV. The experimental data exhibit a slight deviation from linearity, which is remarkably reproduced by the FEM simulation. The best fit is obtained for $\alpha= 0.44 \cdot 10^{23}$ cm$^{-3}$. In the case of H implantations at 2 MeV and 3 MeV, shown in Fig. 4b and 4c, respectively, the trend is linear, indicating that damage densities are well below saturation.

Simulations also produce linear behaviours, and best fits are obtained for $\alpha = 0.88 \cdot 10^{23}$ cm$^{-3}$ (2 MeV) and $\alpha = 1.65 \cdot 10^{23}$ cm$^{-3}$ (3 MeV), respectively.

As mentioned above, $\alpha$ can be regarded as an effective parameter expressing the resilience of the diamond structure to induced damage, i.e. lower values of $\alpha$ indicate a more rapid variation of the material density when subjected to structural damage. Despite the relative simplicity of the above-described phenomenological model and the remaining uncertainty on the value of $\rho_{aC}$ (which will be addressed in more details in future works), the value of $\alpha$ exhibits a systematic variation as a function of ion penetration depth $z$: $\alpha = 0.44 \cdot 10^{23}$ cm$^{-3}$, $0.88 \cdot 10^{23}$ cm$^{-3}$ and $1.65 \cdot 10^{23}$ cm$^{-3}$ and $z\sim 3$ μm, ~25 μm and ~50 μm, for 1.8 MeV He, 2 MeV H and 3 MeV H implantations, respectively. Our result is consistent with the experimental observation that higher damage densities are needed to graphitize diamond at increasing depth, as resulting from several works reported in literature [9; 13; 14; 15].

This can be explained if it is considered that strong internal pressure fields arise in deeper implantations from the rigid diamond matrix surrounding the implanted regions, which could effectively increase the resilience of the structure to progressive amorphization and subsequent graphitization (in the case of post-implantation annealing treatment). Qualitative evidence of this mechanism has been identified in several works, such as [8], while in the present work the mechanism is elucidated in quantitative terms, although by adopting a relatively simple model. In particular, it is worth stressing that in this study a constant value of $\alpha$ was set in the FEM simulation of the effects of each single implantation, but the systematic variation of such parameter for implantation processes characterized by different ion penetration depths indicates that $\alpha$ should indeed be regarded as a mechanical parameter that depends from the depth in the material, where different internal pressure fields are developed. Therefore, the different estimations of $\alpha$ presented here are to be considered as effective values arising from average estimations through each implantation profile.

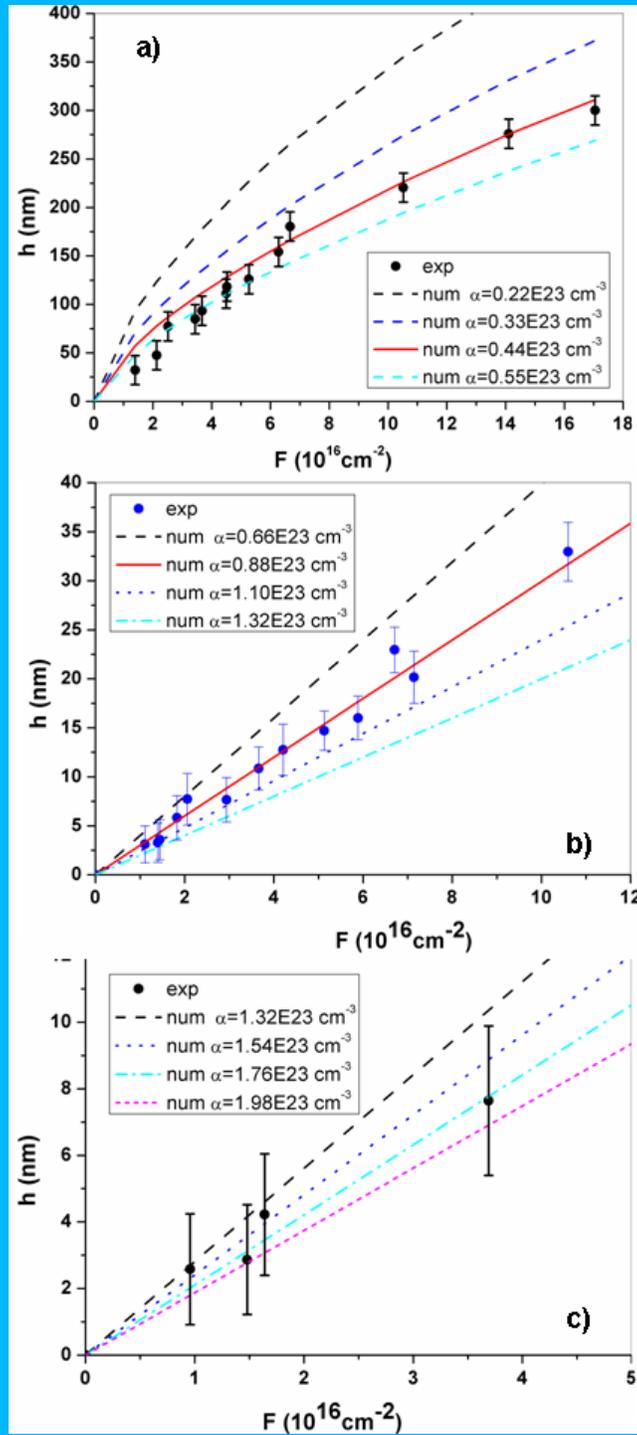

*Fig 4:* Experimental ("exp") and numerical ("num") swelling values *h* vs. fluence *F* for *a)* 1.8 MeV He ions, *b)* 2MeV H ions, and *c*) 3MeV H ions. Error bars are included for experimental points and numerical values are plotted for various values of the saturation density *α*.

## 4. Conclusions

Our modelling of surface swelling in ion-implanted diamond, carried out for the first time using FEM simulations, yields good accordance between experimental and numerical data. In particular, the systematic variation of empirical parameter $\alpha$ for MeV ion implantation at increasing depth accounts for a phenomenon that was observed qualitatively in several previous reports [9; 13; 14; 15], i.e. the progressive resilience of the diamond lattice to graphitization at increasing implantation depth.

The results reported in this paper prove that the FEM numerical analysis can significantly contribute to the quantitative interpretation of the structural damage mechanisms in ion-implanted diamond, and provides encouraging insight for further in-depth analysis and systematic studies at higher implantation fluences, where saturation of swelling values is achieved.

## Acknowledgements

This work is supported by the "Accademia Nazionale dei Lincei – Compagnia di San Paolo" Nanotechnology grant and by "DANTE" and "FARE" experiments of "Istituto Nazionale di Fisica Nucleare" (INFN), which are gratefully acknowledged.